\documentclass [12pt] {report}
\usepackage[cp1251]{inputenc}
\usepackage[dvips]{graphicx}
\usepackage{epsfig, latexsym,graphicx}
\usepackage{psfrag}
\usepackage{amsfonts,amsmath,amssymb,amscd,array, euscript}
\newcommand {\D}[2] {\displaystyle\frac{\partial{#1}}{\partial{#2}}}

\newcommand {\la} {\lambda}
\newcommand {\La} {\Lambda}

\newcommand {\Si} {\Sigma}

\newcommand {\De} {\Delta}

\newcommand {\fr} {\displaystyle\frac}

\newcommand {\be} {\begin{equation}}
\newcommand {\ee} {\end{equation}}
\newcommand {\ba} {\begin{array}}
\newcommand {\ea} {\end{array}}
\newcommand {\bp} {\begin{picture}}
\newcommand {\ep} {\end{picture}}
\newcommand {\bc} {\begin{center}}
\newcommand {\ec} {\end{center}}

\newcommand {\bt} {\begin{tabular}}
\newcommand {\et} {\end{tabular}}
\newcommand {\lf} {\left}
\newcommand {\rg} {\right}

\newcommand {\nin}{\noindent}

\newcommand {\cA} {{\cal A}}

\newcommand {\cF} {{\cal F}}
\newcommand {\cH} {{\cal H}}

\newcommand {\bv} {{\bf v}}

\newcommand {\ses} {\medskip}

\def\2#1#2#3{{#1}_{#2}\hspace{0pt}^{#3}}
\def\3#1#2#3#4{{#1}_{#2}\hspace{0pt}^{#3}\hspace{0pt}_{#4}}
\newcounter{sctn}
\def\sec#1.#2\par{\setcounter{sctn}{#1}\setcounter{equation}{0}
                  \noindent{\bf\boldmath#1.#2}\bigskip\par}
\begin {document}

\begin{titlepage}

\begin{center}
{\Large \bf
  Finslerian grounds for four--directional anisotropic
 \ses

kinematics

 }

\end{center}

\vspace{0.3in}

\begin{center}

\vspace{.15in} {\large G. S. Asanov\\} \vspace{.25in}
{\it Division of Theoretical Physics, Moscow State University\\
119992 Moscow, Russia\\
(e-mail: asanov@newmail.ru)} \vspace{.05in}

\end{center}

\begin{abstract}

Upon straightforward four--directional  extension of
the special--relativistic  two--dimensional transformations to the four--dimensional case
we lead to convenient totally anisotropic kinematic transformations,
which prove to reveal many remarkable group and invariance
properties. Such a promise is shown to ground the  basic  manifold
with the Finslerian fourth-root metric function to measure length of
relativistic four--vectors. Conversion to the framework of
relativistic four--momentum is also elucidated. The relativity
principle is strictly retained. An interesting particular algebra
for subtraction and composition of three-dimensional
relative velocities is arisen.
The correspondence principle is
operative in the sense that at small relative velocities the
transformations introduced tend approximately to ordinary Lorentzian
precursors. The transport synchronization remains valid.

Abbreviation RF will be used for (inertial) reference frames.


\end{abstract}

\end{titlepage}

\vskip 1cm

\setcounter{sctn}{1} \setcounter{equation}{0}

{\bf 1. Introduction and Motivation }

\bigskip

The present work is inspired by an intriguing  possibility (previously opened in [1]) to extend the ordinary
Lorentzian transformations
by suppressing any isotropy assumption.
It is rather astonishing fact traced below
that if we introduce the kinematic transformations in such a four--directional way
 (eqs. (2.1)--(2.5) below),
 we find the geometric
grounds to be given exactly by the fourth--root Finslerian metric function served to measure the length of four--vectors
(instead of  the pseudoeuclidean quadratic method).
This way gives rise to a totally anisotropic and also  geometrically motivated  relativistic
framework, which is self--consistent.
The relevant usage of the Finsler geometry methods [2,3] in future seems to be indispensable.

Generally,  the availability of totally isotropic alternative kinematics may be of help
 to provide theoretical instruments to analyze real
 experimental limits for isotropy rooted in the Lorentzian relativistic framework.
There  may also exist situations in Universe or in high--energy particle interactions
where anisotropy may be a reality (relevant discussions can be found in [4]).
Various possibilities may occur in future experiments to face with space--time anisotropies.
At least, the conventional isotropy of space--time may be ``not full truth'' even at the kinematical level.
In any case, the very desire to have rigorous anisotripic relativistic framework motivated geometrically
to judge upon possible relativistic anisotropies is justified.

Our consideration developed below  is founded on the assumption that  the four geometrically distinguished directions exist
in  the space--time domain to which the consideration refers, provided that there is no preference among the directions.
We commence with Section 2, where
our decisive step is to ``enlarge''  the ordinary special-relativistic Lorentzian transformations
$
Y'^0= (Y^0+s^1Y^1) /\sqrt{(1+s^1)(1-s^1)},
~
Y'^1= (s^1Y^0+Y^1)/ {\sqrt{(1+s^1)(1-s^1)}}
$
 in a
straightforward way from the dimension 2 to the dimension 4.
The choice (2.1)--(2.5) made under the four--directional breakdown of space isotropy
 entails numerous remarkable implications.

Namely, the invoked kinematic coefficients, like   their Lorentzian precursors, are symmetric and
of the unit determinant.
 The Lorentzian square root is replaced by  the fourth--root
function $
A(s)
$
given by (2.5). Remarkably, the fourth--degree
Finslerian metric function, to be denoted by $\cF$,  can be found upon stipulating that the function
is left invariant under the kinematic transformations adopted.
 Transition to the respective $\cA^{\{+\}}_4$--co--treatment is
straightforward, so that the forms of the associated Hamiltonian function and the dispersion relation,
the
Finslerian
transformation laws for momenta and frequencies,
as well as due Finslerian corrections to Doppler effect can unambiguously and explicitly be obtaining.

Non--trivial  expressions can be explicated for the  inversion, subtraction, and composition of relative three--dimensional
 velocities
(Section 3).
 The ordinary reciprocity principle
(which claims that  given a three-dimensional velocity $\bv$ the reciprocal velocity is $-\bv$)
fails. Instead, an extended inversion operation $\ominus$ is to be substituted with the ordinary minus.
In particular, the dependence of the momentum on velocities remarkably follows such a rule (according to (2.43)).
The $\cA^{\{+\}}_4$--composition law is symmetric, while the $\cA^{\{+\}}_4$--subtraction law
proves to be  antisymmetric in only the sense assigned
by the operation $\ominus$. The respective algebra  fulfills the associative rule. Four invariant vectors are arisen.
The abelian group nature
 is inherited by
the $\cA^{\{+\}}_4$--kinematic transformations  from the ordinary two--dimensional
special--relativistic Lorentz transformation.

Despite the circumstance that the kinematic transformations (2.1)--(2.4)
do deviate drastically from the Lorentzian case proper,
 nevertheless they fulfill the conventional correspondence principle
in a rather strong way. Indeed, the law--velocity approximations (Section 4) show
% that holds in the $\cA^{\{+\}}_4$--framework, namely
 that the post--Lorentzian corrections to the fundamental
fourth root function $A(s)$ (see (2.5)), and hence to
   the very $\cA^{\{+\}}_4$--kinematic transformations (2.1)--(2.4),
start deviating from their ordinary Lorentzian precursors with but
the O(3)--terms (quite similar conclusion can be addressed to the
dependence of the particle energy on the three--dimensional momentum).
The subtraction and composition laws for velocities reveal the O(2)--order corrections.

The paper ends with  Conclusions in which we shortly discuss the key aspects of our approach.

\ses\ses

\setcounter{sctn}{2} \setcounter{equation}{0}

{\bf 2. Principal observations
}

\bigskip

Let us introduce the
$\cA_4^{\{+\}}$--{\it kinematic transformations}
\be
Y'^0=\fr
{Y^0+s^1Y^1+s^2Y^2+s^3Y^3}
{A(s)},
\ee
\ses
\be
Y'^1=\fr
{s^1Y^0+Y^1+s^3Y^2+s^2Y^3}
{A(s)},
\ee
\ses
\be
Y'^2=\fr
{s^2Y^0+s^3Y^1+Y^2+s^1Y^3}
{A(s)},
\ee
\ses
\be
Y'^3=\fr
{s^3Y^0+s^2Y^1+s^1Y^2+Y^3}
{A(s)}
\ee
with
\be
A(s)=\sqrt[4]{(1+s^1+s^2+s^3)(1-s^1+s^2-s^3)(1+s^1-s^2-s^3)(1-s^1-s^2+s^3)}
\ee
to describe the transition from a RF $\Si=\Si(s)$ into another RF $\Si'$, so that
$\{Y^p\}\in \Si$ and $\{Y'^p\}\in \Si'$. Here,
 $s=\{s^a\}=\{s^1,s^2,s^3\}$ are components of  three--dimensional motion velocity
of  $\Si$ with respect to $\Si'$.
The indices $a,b,...$ , $p,q,...$, and $A,B$ will respectively run over
the range 1,2,3, the range  0,1,2,3, and the range 1,2,3,4.
The designation
$\cA_4^{\{+\}}$ means that the approach is anisotropic, the dimension is 4, and  the positivity conditions of the types
\be
1+s^1+s^2+s^3>0,~1-s^1+s^2-s^3>0,~1+s^1-s^2-s^3>0,~1-s^1-s^2+s^3>0
\ee
and
\be
Y^0\!+\!Y^1\!+\!Y^2\!+\!Y^3>0,~
Y^0\!-\!Y^1\!+\!Y^2\!-\!Y^3>0,~
Y^0\!+\!Y^1\!-\!Y^2\!-\!Y^3>0,~
Y^0\!-\!Y^1\!-\!Y^2\!+\!Y^3>0
\ee
are rigorously fulfilled.

%\pgbrk

If we write the transformations in the vector form
\be
 Y'^p=\La^p_q(s)Y^q,
\ee
the kinematic coefficients involve the components
\be
\La_0^0(s)=\La_1^1(s)=\La_2^2(s)=\La_3^3(s)=\fr1{A(s)},
\ee
\ses
\be
\La_0^a(s)=\La_a^0(s)=\fr1{A(s)}s^a\equiv \La_0^0(s)s^a,
\ee
\ses
\be
\La_1^2(s)=\La_2^1(s)=\fr1{A(s)}s^3,\quad \La_1^3(s)=\La_3^1(s)=\fr1{A(s)}s^2,\quad \La_2^3(s)=\La_3^2(s)=\fr1{A(s)}s^1.
\ee
The symmetry
\be
\La^p_q(s)=\La^q_p(s)
 \ee
 and the determinant unity
\be
 \det(\La^p_q)(s)=1
 \ee
hold fine at any value of the set $s^a=\{s^1,s^2,s^3\}$.

%\pgbrk

The  $\cA^{\{+\}}_4$--{\it kinematic length}
%$\cF(Y)$ of the vector  reads
\be
\cF(Y)=
\sqrt[4]{(Y^0\!+\!Y^1\!+\!Y^2\!+\!Y^3)
(Y^0\!-\!Y^1\!+\!Y^2\!-\!Y^3)
(Y^0\!+\!Y^1\!-\!Y^2\!-\!Y^3)
(Y^0\!-\!Y^1\!-\!Y^2\!+\!Y^3)}
\ee
of the vector $\{Y^p\}$  fulfills
the $\cA^{\{+\}}_4$--{\it kinematic invariance}
\be
\cF(Y')=\cF(Y).
\ee

There exists a simple way to verify the  invariance.
Namely,
applying the coefficients (2.9)--(2.11) to the parentheses appeared under the
root in the right--hand part of (2.14) yields
\be
Y^0\!+\!Y^1\!+\!Y^2\!+\!Y^3 =\la_1
%\fr{(1+s^1+s^2+s^3)}{A(s)}
(Y^0\!+\!Y^1\!+\!Y^2\!+\!Y^3),
\ee
\ses
\be
Y^0\!-\!Y^1\!+\!Y^2\!-\!Y^3=\la_2
%\fr{(1-s^1+s^2-s^3)}{A(s)}
(Y^0\!-\!Y^1\!+\!Y^2\!-\!Y^3),
\ee
\ses
\be
Y^0\!+\!Y^1\!-\!Y^2\!-\!Y^3=\la_3
%\fr{(1+s^1-s^2-s^3)}{A(s)}
(Y^0\!+\!Y^1\!-\!Y^2\!-\!Y^3),
\ee
\ses
\be
Y^0\!-\!Y^1\!-\!Y^2\!+\!Y^3=\la_4
%\fr{(1-s^1-s^2+s^3)}{A(s)}
(Y^0\!-\!Y^1\!-\!Y^2\!+\!Y^3),
\ee
where
\be
\la_1=\fr{1\!+\!s^1\!+\!s^2\!+\!s^3}{A(s)},~\la_2=\fr{1\!-\!s^1\!+\!s^2\!-\!s^3}{A(s)},
~\la_3=\fr{1\!+\!s^1\!-\!s^2\!-\!s^3}{A(s)}, ~\la_4=\fr{1\!-\!s^1\!-\!s^2\!+\!s^3}{A(s)},
\ee
so that  on taking into account the right--hand part in the expression
(2.5)
of  $ A(s) $
we may just establish (2.15)
(that is to say, the quantities $s^a$ do  disappear in the right--hand part of (2.14)).

%\pgbrk

The  {\it isotropic four--vectors} $C_A=\{C_A^p\}$
with
\be
C_1^0=-C_1^1-C_1^2-C_1^3,
\quad
C_2^0=C_2^1-C_2^2+C_2^3,
\quad
C_3^0=-C_3^1+C_3^2+C_3^3,
\quad
C_4^0=C_4^1+C_4^2-C_4^3
\ee
show the property
\be
\cF(C_A)=0.
\ee

%\pgbrk

Also, the
$\cA_4^{\{+\}}$-- {\it momentum  kinematic transformations}
\be
P'_0=\fr
{P_0+f^1P_1+f^2P_2+f^3P_3}
{A(f)},
\ee
\ses
\be
P'_1=\fr
{f^1P_0+P_1+f^3P_2+f^2P_3}
{A(f)},
\ee
\ses
\be
P'_2=\fr
{f^2P_0+f^3P_1+P_2+f^1P_3}
{A(f)},
\ee
\ses
\be
P'_3=\fr
{f^3P_0+f^2P_1+f^1P_2+P_3}
{A(f)}
\ee
can be found, where
\be
A(f)=\sqrt[4]{(1+f^1+f^2+f^3)(1-f^1+f^2-f^3)(1+f^1-f^2-f^3)(1-f^1-f^2+f^3)}
\ee
with
\be
 f=\ominus s
\ee
(here the right--hand part is the operation given by the formulae (3.1)--(3.4) of the next section)
so that the contractions of co-- and contra--vectors remain invariant under the transformations considered, that is,
\be
P'_qY'^q=P_qY^q.
\ee
The associated
$\cA_4^{\{+\}}$--{\it Hamiltonian function}
\be
\cH(P)=
\sqrt[4]{(P_0\!+\!P_1\!+\!P_2\!+\!P_3)
(P_0\!-\!P_1\!+\!P_2\!-\!P_3)
(P_0\!+\!P_1\!-\!P_2\!-\!P_3)
(P_0\!-\!P_1\!-\!P_2\!+\!P_3)}
\ee
gives rise to the
 $\cA_4^{\{+\}}$--{\it dispersion relation}
\be
E=E(m;P_1,P_2,P_3)
\ee
which explicit form is the following:
\be
(E+ P_1+P_2+P_3)
(E-P_1+P_2-P_3)
(E+P_1-P_2-P_3)
(E-P_1-P_2+P_3)
=m^4,
\ee
where $m$ is the particle mass. Given the momentum components $\{P_1,P_2,P_3\}$,
   from (2.32)  energy $E$ of the particle is found.

Similarly to (2.15), the Hamiltonian  function (2.30) is $\cA_4^{\{+\}}$--invariant:
\be
\cH(P')=\cH(P).
\ee

%\pgbrk

Applying the Finslerian rule
\be
Y_p=\fr12\D{(\cF(Y))^2}{Y^p}
\ee
(see [1--3]) to introduce covariant vectors, from (2.14) we obtain for a single particle
the $\cA_4^{\{+\}}$--{\it four--momentum}
$P_p=P_p(m;V)$ with  the explicit components
\be
\fr{P_0}{m\cF(V)}=
\fr{1}{V^0\!+\!V^1\!+\!V^2\!+\!V^3}
+\fr{1}{V^0\!-\!V^1\!+\!V^2\!-\!V^3}
+\fr{1}{V^0\!+\!V^1\!-\!V^2\!-\!V^3}
+\fr{1}{V^0\!-\!V^1\!-\!V^2\!+\!V^3},
\ee
\ses
\be
\fr{P_1}{m\cF(V)}=
\fr{1}{V^0\!+\!V^1\!+\!V^2\!+\!V^3}
-\fr{1}{V^0\!-\!V^1\!+\!V^2\!-\!V^3}
+\fr{1}{V^0\!+\!V^1\!-\!V^2\!-\!V^3}
-\fr{1}{V^0\!-\!V^1\!-\!V^2\!+\!V^3},
\ee
\ses
\be
\fr{P_2}{m\cF(V)}=
\fr{1}{V^0\!+\!V^1\!+\!V^2\!+\!V^3}
+\fr{1}{V^0\!-\!V^1\!+\!V^2\!-\!V^3}
-\fr{1}{V^0\!+\!V^1\!-\!V^2\!-\!V^3}
-\fr{1}{V^0\!-\!V^1\!-\!V^2\!+\!V^3},
\ee
\ses
\be
\fr{P_3}{m\cF(V)}=
\fr{1}{V^0\!+\!V^1\!+\!V^2\!+\!V^3}
-\fr{1}{V^0\!-\!V^1\!+\!V^2\!-\!V^3}
-\fr{1}{V^0\!+\!V^1\!-\!V^2\!-\!V^3}
+\fr{1}{V^0\!-\!V^1\!-\!V^2\!+\!V^3}.
\ee
The four--vector $V^p$ entered represents the four--velocity of particle.
With the formulae (2.14), (2.30), and (2.35)--(2.38)
it can straightforwardly be verified that
\be
\cH(P)=m.
\ee
The right--hand parts in (2.35)--(2.38)
 can be simplified,  yielding the representation
\be
P_0=m\fr{(V^0)^3-V^0\lf((V^1)^2+(V^2)^2+(V^3)^2\rg)+2V^1V^2V^3}
{\Bigl(
(V^0\!+\!V^1\!+\!V^2\!+\!V^3)
(V^0\!-\!V^1\!+\!V^2\!-\!V^3)
(V^0\!+\!V^1\!-\!V^2\!-\!V^3)
(V^0\!-\!V^1\!-\!V^2\!+\!V^3)
\Bigr)^{3/4}
}
%\lf(\cF(V)\rg)^3}
\ee
and, in terms of the three-dimensional vectors
\be
v^a=V^a/V^0
\ee
and
\be
 p_a=P_a/P_0,
\ee
the equality
\be
\fr1m p_a=\ominus(v^a),
\ee
where $\ominus$ is the operation presented in the next section by means of the formulae
(3.1)--(3.4).

\ses\ses

\setcounter{sctn}{3} \setcounter{equation}{0}

{\bf 3. Subtraction and composition laws for velocities
}

\bigskip

First of all, if we consider the inverse transformations (from the RF
$\Si'$ into the RF $\Si$), we find the  $\cA_4^{\{+\}}$--{\it reciprocal velocity}
$\ominus s$
with the following components
\be
(\ominus s)^1=
-
\fr1{K(s)}
%\fr1{(A(s))^4}
\Biggl[s^1-2s^2s^3-(s^1)^3+s^1(s^2s^2+s^3s^3)\Biggr],
\ee
%\ses
\be
(\ominus s)^2=
-
\fr1{K(s)}
%\fr1{(A(s))^4}
\Biggl[s^2-2s^1s^3-(s^2)^3+s^2(s^1s^1+s^3s^3)\Biggr],
\ee
%\ses
\be
(\ominus s)^3=
-
\fr1{K(s)}
%\fr1{(A(s))^4}
\Biggl[s^3-2s^1s^2-(s^3)^3+s^3(s^1s^1+s^2s^2)\Biggr],
\ee
where
\be
K(s)=1-s^1s^1-s^2s^2-s^3s^3+2s^1s^2s^3.
\ee

%\pgbrk

Also, by
considering a succession of three RFs $\Si,\Si',\Si''$ and denoting by
$ s_{\{1\}},~s_{\{2\}},~s_{\{3\}} $ their respective mutual motion velocities,
%$ s^a_{\{1\}}=s^a_{\{1\}}(a,b),~s^a_{\{2\}}=s^a_{\{2\}}(b,c),~s^a_{\{3\}}=s^a_{\{3\}}(a,c) $
we can derive from (2.1)--(2.4) the
%For the velocity,  we obtain the extended composition law as well the subtraction law in explicit forms.
$\cA_4^{\{+\}}$--{\it  subtraction  law}
\be
s_{\{1\}}=s_{\{3\}}\ominus s_{\{2\}}
\ee
with the following component dependencies:
\be
Hs^1_{\{1\}}\!=\!
\fr{1\!+\!s^1_{\{3\}}\!+\!s^2_{\{3\}}\!+\!s^3_{\{3\}}}
{1\!+\!s^1_{\{2\}}\!+\!s^2_{\{2\}}+\!s^3_{\{2\}}}
-
\fr{1\!-\!s^1_{\{3\}}\!+\!s^2_{\{3\}}\!-\!s^3_{\{3\}}}
{1\!-\!s^1_{\{2\}}\!+\!s^2_{\{2\}}\!-\!s^3_{\{2\}}}
+
\fr{1\!+\!s^1_{\{3\}}\!-\!s^2_{\{3\}}\!-\!s^3_{\{3\}}}
{1\!+\!s^1_{\{2\}}\!-\!s^2_{\{2\}}\!-\!s^3_{\{2\}}}
-\fr
{1\!-\!s^1_{\{3\}}\!-\!s^2_{\{3\}}\!+\!s^3_{\{3\}}}
{1\!-\!s^1_{\{2\}}\!-\!s^2_{\{2\}}\!+\!s^3_{\{2\}}},
\ee
\ses
\be
Hs^2_{\{1\}}\!=\!
\fr{1\!+\!s^1_{\{3\}}\!+\!s^2_{\{3\}}\!+\!s^3_{\{3\}}}
{1\!+\!s^1_{\{2\}}\!+\!s^2_{\{2\}}+\!s^3_{\{2\}}}
+
\fr{1\!-\!s^1_{\{3\}}\!+\!s^2_{\{3\}}\!-\!s^3_{\{3\}}}
{1\!-\!s^1_{\{2\}}\!+\!s^2_{\{2\}}\!-\!s^3_{\{2\}}}
-
\fr{1\!+\!s^1_{\{3\}}\!-\!s^2_{\{3\}}\!-\!s^3_{\{3\}}}
{1\!+\!s^1_{\{2\}}\!-\!s^2_{\{2\}}\!-\!s^3_{\{2\}}}
-\fr
{1\!-\!s^1_{\{3\}}\!-\!s^2_{\{3\}}\!+\!s^3_{\{3\}}}
{1\!-\!s^1_{\{2\}}\!-\!s^2_{\{2\}}\!+\!s^3_{\{2\}}},
\ee
\ses
\be
Hs^3_{\{1\}}\!=\!
\fr{1\!+\!s^1_{\{3\}}\!+\!s^2_{\{3\}}\!+\!s^3_{\{3\}}}
{1\!+\!s^1_{\{2\}}\!+\!s^2_{\{2\}}+\!s^3_{\{2\}}}
-
\fr{1\!-\!s^1_{\{3\}}\!+\!s^2_{\{3\}}\!-\!s^3_{\{3\}}}
{1\!-\!s^1_{\{2\}}\!+\!s^2_{\{2\}}\!-\!s^3_{\{2\}}}
-
\fr{1\!+\!s^1_{\{3\}}\!-\!s^2_{\{3\}}\!-\!s^3_{\{3\}}}
{1\!+\!s^1_{\{2\}}\!-\!s^2_{\{2\}}\!-\!s^3_{\{2\}}}
+\fr
{1\!-\!s^1_{\{3\}}\!-\!s^2_{\{3\}}\!+\!s^3_{\{3\}}}
{1\!-\!s^1_{\{2\}}\!-\!s^2_{\{2\}}\!+\!s^3_{\{2\}}},
\ee
\ses\ses
where
\be
H\!=\!
\fr{1\!+\!s^1_{\{3\}}\!+\!s^2_{\{3\}}\!+\!s^3_{\{3\}}}
{1\!+\!s^1_{\{2\}}\!+\!s^2_{\{2\}}+\!s^3_{\{2\}}}
+
\fr{1\!-\!s^1_{\{3\}}\!+\!s^2_{\{3\}}\!-\!s^3_{\{3\}}}
{1\!-\!s^1_{\{2\}}\!+\!s^2_{\{2\}}\!-\!s^3_{\{2\}}}
+
\fr{1\!+\!s^1_{\{3\}}\!-\!s^2_{\{3\}}\!-\!s^3_{\{3\}}}
{1\!+\!s^1_{\{2\}}\!-\!s^2_{\{2\}}\!-\!s^3_{\{2\}}}
+\fr
{1\!-\!s^1_{\{3\}}\!-\!s^2_{\{3\}}\!+\!s^3_{\{3\}}}
{1\!-\!s^1_{\{2\}}\!-\!s^2_{\{2\}}\!+\!s^3_{\{2\}}}.
\ee

%\pgbrk

The $\cA_4^{\{+\}}$--{\it composition law}
\be
s_{\{3\}}=s_{\{1\}}\oplus s_{\{2\}}
\ee
can be evaluated in a similar fashion, yielding
\be
s^1_{\{3\}}=
\fr
{s_{\{1\}}^1+s_{\{2\}}^1+s_{\{1\}}^2s_{\{2\}}^3+s_{\{1\}}^3s_{\{2\}}^2}
{1+s_{\{1\}}^1s_{\{2\}}^1
+s_{\{1\}}^2s_{\{2\}}^2
+s_{\{1\}}^3s_{\{2\}}^3},
\ee
\ses
\be
s^2_{\{3\}}=
\fr
{s_{\{1\}}^2
+s_{\{2\}}^2+s_{\{1\}}^1s_{\{2\}}^3+s_{\{1\}}^3s_{\{2\}}^1}
{1+s_{\{1\}}^1s_{\{2\}}^1
+s_{\{1\}}^2s_{\{2\}}^2+s_{\{1\}}^3s_{\{2\}}^3},
\ee
\ses
\be
s^3_{\{3\}}=
\fr
{s_{\{1\}}^3+s_{\{2\}}^3+s_{\{1\}}^1s_{\{2\}}^2+s_{\{1\}}^2s_{\{2\}}^1}
{1+s_{\{1\}}^1s_{\{2\}}^1+s_{\{1\}}^2s_{\{2\}}^2+s_{\{1\}}^3s_{\{2\}}^3}.
\ee
An alternative convenient way to obtain the formulae (3.11)--(3.13) is to resolve the subtraction law  set
(3.6)--(3.9)
with respect to the  quantities
$s^1_{\{3\}},\,s^2_{\{3\}},\,s^3_{\{3\}}$.

%\pgbrk

The  $\cA^{\{+\}}_4$--algebra thus arisen to operate with three--dimensional velocities
is symmetric
\be
s_{\{1\}}\oplus s_{\{2\}}=
s_{\{2\}}\oplus s_{\{1\}}
\ee
and
also antisymmetric in terms of the inversion operator $\ominus$:
\be.
s_{\{1\}}\ominus s_{\{2\}}=\ominus\Bigl(s_{\{2\}}\ominus s_{\{1\}}\Bigr).
\ee
The associative law
\be
\Bigl(s_{\{1\}}\oplus s_{\{2\}}\Bigr)\oplus s_{\{3\}}
=
s_{\{1\}}\oplus\Bigl( s_{\{2\}}\oplus s_{\{3\}}\Bigr)
\ee
holds.
The identities
\be
s\ominus  s=0,
\ee
\ses
\be
0\ominus  s=\ominus s,
\ee
\ses
\be
0\oplus  s= s,
\ee
\ses
\be
s\oplus \Bigl(\ominus s\Bigr)=0,
\ee
\ses
\be
s_{\{1\}}\ominus\Bigl(\ominus s_{\{2\}}\Bigr)=s_{\{1\}}\oplus s_{\{2\}},
\ee
\ses
\be
s_{\{1\}}\ominus s_{\{2\}}\oplus s_{\{2\}}\oplus s_{\{3\}}=
s_{\{1\}}\oplus s_{\{3\}},
\ee
\ses
\be
s_{\{1\}}\oplus\Bigl(\ominus s_{\{2\}}\Bigr)=s_{\{1\}}\ominus s_{\{2\}}
\ee
can straightforwardly be verified.
Symbolically, we can write
\be
\ominus\ominus=\oplus
\ee
and
\be
\oplus\ominus=\ominus\oplus=\ominus.
\ee
% $ c_A=I_A^a/I_A^0$
The definitions
 \be
c_1=\{-1,-1,-1\},\quad c_2=\{1,-1,1\},\quad c_3=\{-1,1,1\},\quad c_4=\{1,1,-1\},
\ee
introduce  four  three-dimensional {\it invariant velocities} $\{c_A\}$, symbolically
\be
\La\cdot  c_A= c_A.
\ee
They reveal the properties
\be
s\oplus c_A=s, \quad
s\ominus c_A=s,
\ee
and
\be
\ominus c_A=-c_A.
\ee

% \pgbrk

The group property
\be
\La^p_q(s_{\{1\}})\La^q_r(s_{\{2\}})=
\La^p_r(s_{\{1\}}\oplus s_{\{2\}})
 \ee
and the symmetry
\be
\La^p_q(s_{\{1\}})\La^q_r(s_{\{2\}})=
\La^p_q(s_{\{2\}})\La^q_r(s_{\{1\}})
 \ee
together with the inverting
 \be
(\La^p_q(s))^{-1}=\La^p_q(\ominus s)
\ee
are valid.

%\pgbrk

In checking the momentum transformations (2.23)--(2.28) it is useful to previously establish the identities
\be
1+(\ominus s)^1+(\ominus s)^2+(\ominus s)^3=\fr1{K(s)}(1-s^2+s^2-s^3)
(1+s^2-s^2-s^3)(1-s^2-s^2+s^3),
\ee
\ses
\be
1-(\ominus s)^1+(\ominus s)^2-(\ominus s)^3=\fr1{K(s)}(1+s^2+s^2+s^3)
(1+s^2-s^2-s^3)(1-s^2-s^2+s^3),
\ee
\ses
\be
1+(\ominus s)^1-(\ominus s)^2-(\ominus s)^3=\fr1{K(s)}(1+s^2+s^2+s^3)
(1-s^2+s^2-s^3)(1-s^2-s^2+s^3),
\ee
\ses
\be
1+(\ominus s)^1-(\ominus s)^2-(\ominus s)^3=\fr1{K(s)}(1+s^2+s^2+s^3)
(1-s^2+s^2-s^3)(1-s^2-s^2+s^3)
\ee
together with
\be
A(\ominus s)=\fr1{K(s)}(A(s))^3.
\ee

\ses\ses

 \setcounter{sctn}{4} \setcounter{equation}{0}

{\bf 4. Approximations }

\bigskip

The transformations (2.1)--(2.5)  obviously extend
 the ordinary special--relativistic Lorentz transformations
\be
Y'^0_{\{\rm special~ Lorentzian\}}=\fr
{Y^0+s^1Y^1}
{\sqrt{(1+s^1)(1-s^1)}}, \qquad
Y'^1_{\{\rm special~ Lorentzian\}}=\fr
{s^1Y^0+Y^1}
{\sqrt{(1+s^1)(1-s^1)}},
\ee
accordingly  the Finslerian result (2.14) extends the pseudoeuclidean function
\be
{\cF(Y)}_{\{\rm special~ Lorentzian\}}=\sqrt{(Y^0+Y^1)(Y^0-Y^1)}.
\ee

 From (3.5)--(3.13) the ordinary two--dimensional special--relativistic formulae ensue
as follows:
\be
\{s_{\{1\}}^2=s_{\{1\}}^3=s_{\{2\}}^2=s_{\{2\}}^3=0\}\to s_{\{1\}}^1=
\fr{s_{\{3\}}^1-s_{\{2\}}^1}{1-s_{\{2\}}^1s_{\{3\}}^1}
\quad
{\text{and}} \quad
s_{\{3\}}^1=\fr{s_{\{1\}}^1+s_{\{2\}}^1}{1+s_{\{1\}}^1s_{\{2\}}^1}.
\ee
In this case, the reciprocity of the ordinary ``obvious''  type
\be
\ominus s^a=-s^a
\ee
is a true implication from the law (3.1)--(3.4).

When
$$
(s^1)^2+(s^2)^2+(s^3)^2 \ll 1,
$$ in the law--velocity approximation up to O(5) we obtain
\be
A(s)\approx
A_1(s)+A_2(s)
\ee
with \be A_1(s)=1
-\fr12\Bigl(
(s^1)^2+(s^2)^2+(s^3)^2\Bigr)
-\fr18\Bigl(
(s^1)^4+(s^2)^4+(s^3)^4\Bigr)
\ee
and
\be
A_2(s)=2s^1s^2s^3
-\fr54\Bigl(
(s^1)^2(s^2)^2+(s^2)^2(s^3)^2+(s^1)^2(s^3)^2\Bigr).
\ee
Inverting yields
\be
\fr1{A(s)}\approx
(A^{-1})_{1}(s)+(A^{-1})_{2}(s)
\ee
with
\be
(A^{-1})_{1}(s)=1
+\fr12\Bigl(
(s^1)^2+(s^2)^2+(s^3)^2\Bigr)
+\fr38\Bigl(
(s^1)^4+(s^2)^4+(s^3)^4\Bigr)
\ee
and
\be
(A^{-1})_{2}(s)=-2s^1s^2s^3
+\fr74\Bigl(
(s^1)^2(s^2)^2+(s^2)^2(s^3)^2+(s^1)^2(s^3)^2\Bigr).
\ee

%\pgbrk

Also, we approximate the reciprocity law
\be
(\ominus s)^1\approx
-s^1-(s^1)^2-(s^2-s^3)^2,
\ee
%\ses
\be
(\ominus s)^2\approx
-
s^2
-(s^2)^2-(s^1-s^3)^2,
\ee
%\ses
\be
(\ominus s)^3\approx
-
s^3
-(s^3)^2 -(s^1-s^2)^2,
\ee
the subtraction law
\be
(s_{\{3\}}\ominus s_{\{2\}})^1
 \approx
 s^1_{\{3\}}-s^1_{\{2\}}
+2 s^2_{\{2\}}s^3_{\{2\}}
-  s^2_{\{2\}}s^3_{\{3\}}
-   s^2_{\{3\}}s^3_{\{2\}},
\ee
\ses
\be
(s_{\{3\}}\ominus s_{\{2\}})^2
 \approx  s^2_{\{3\}}-s^2_{\{2\}}
+2 s^1_{\{2\}}s^3_{\{2\}}
-  s^1_{\{2\}}s^3_{\{3\}}
-   s^1_{\{3\}}s^3_{\{2\}},
\ee
\ses
\be
(s_{\{3\}}\ominus s_{\{2\}})^3
\approx  s^3_{\{3\}}-s^3_{\{2\}}
+2 s^1_{\{2\}}s^2_{\{2\}}
-  s^1_{\{2\}}s^2_{\{3\}}
-   s^1_{\{3\}}s^2_{\{2\}},
\ee
and the composition law
\be
(s_{\{1\}}\oplus s_{\{2\}})^1
\approx
 s^1_{\{2\}}+s^1_{\{3\}}
+
s^2_{\{2\}}s^3_{\{3\}}
+s^3_{\{2\}}s^2_{\{3\}},
\ee
\ses
\be
(s_{\{1\}}\oplus s_{\{2\}})^2
\approx  s^2_{\{2\}}+s^2_{\{3\}}
+
s^1_{\{2\}}s^3_{\{3\}}
+s^3_{\{2\}}s^1_{\{3\}},
\ee
\ses
\be
(s_{\{1\}}\oplus s_{\{2\}})^3
\approx  s^3_{\{2\}}+s^3_{\{3\}}
+
s^1_{\{2\}}s^2_{\{3\}}
+s^2_{\{2\}}s^1_{\{3\}}.
\ee

%\pgbrk

The law--momentum approximation for the dispersion relation (2.32) starts with
\be
E\approx
m+\fr1{2m}\Bigl(
(P_1)^2+(P_2)^2+(P_3)^2\Bigr)
+\fr1{8m^3}\Bigl(
(P_1)^4+(P_2)^4+(P_3)^4\Bigr)+E_{\cA},
\ee
where
\be
E_{\cA}\approx -\fr2{m^2}P_1P_2P_3
-\fr5{4m^3}\Bigl(
(P_1)^2(P_2)^2+(P_2)^2(P_3)^2+(P_1)^2(P_3)^2\Bigr)
\ee
is
the $\cA_4^{\{+\}}$--correction.

\ses\ses

{\bf Conclusions }

\bigskip

Geometrically, consideration in the present work was  founded  upon the relativistic and geometric properties of
the space--time endowed tentatively with the four--directional Finslerian metric function (2.14)
(the uni--directional anisotropic relativity was developed in [5--8] in context of the pseudo--Finsleroid approach).
Neither stipulating  existence of a preferred reference frame nor assuming violations of the relativity principle
have been implied.
An important common feature of the ordinary
pseudoeuclidean theory of special relativity
and of the Finslerian relativistic approach
is that
they both endeavor to establish a universal prescription
for applying the theory to systems in differing states of motion.

In  the $\cA_4^{\{+\}}$--approach, the { \it slow transport synchronization is fulfilled exactly},
which  is, in fact, a direct implication
of the linear--nature of lowest--order approximation of the $\cA_4^{\{+\}}$--composition law for velocities
(see  (4.17)--(4.19)).
A convenient more detailed motivation thereto is the following.
If a clock is considered to move in the  RF $\Si=\Si(s)$ with a velocity $u^a$, so that
$\De Y^a=u^a\De Y^0$, then with respect to the  RF $\Si'$ the clock velocity is $w^a=u^a\oplus s^a$ and we obtain
the value $\De Y^0_{{\rm clock}}=A(w)\De Y'^0$ for the time which the moving clock shows in its rest frame.
Using here (2.1) and noting the expansions (4.5)--(4.7) and (4.17)--(4.19), we just
arrive at the conclusion that
$\lf(\partial (\De Y^0_{{\rm clock}}/\De Y^0)/\partial s^a\rg)|_{s^a=0}=0$.
The conclusion is tantamount to the meaning of the slow transport synchronization.

%The occurrence of  four invariant velocities $c_A$ (given by (3.26)) extends the ordinary role of
%the light velocities and challenge to creating a relevant $\cA_4^{\{+\}}$--electrodynamics.
%We postpone  analyzing such aspects
%formulating light--behaviour aspects of the $\cA_4^{\{+\}}$--approach
%to special papers.

Generally,
the theoretical structure and
experimental limitations of the special relativity
cannot be tested in all its aspects
unless
a self-consistent alternative metric theory
is applied.
We expect that it is the Finsler-type geometries that
may born such  theories,
advancing simultaneously {\it
new
philosophic-physical sights
on the relationship between Relativity and Geometry}.

\ses \ses

\def\bibit[#1]#2\par{\rm\noindent\parskip1pt
                     \parbox[t]{.05\textwidth}{\mbox{}\hfill[#1]}\hfill
                     \parbox[t]{.925\textwidth}{\baselineskip11pt#2}\par}

\nin {\bf References}
\bigskip

\bibit[1] G. S. Asanov:  Finslerian metric function of totally anisotropic type.
 Relativistic aspects,
 arXiv:math-ph/0510007.

\bibit[2] G. S. Asanov: \it Finsler Geometry, Relativity and Gauge Theories, \rm
D.~Reidel Publ. Comp., Dordrecht 1985.

\bibit[3] H.~ Rund: \it The Differential Geometry of Finsler spaces, \rm
Springer-Verlag, Berlin 1959.

\bibit[4] I. Zborovsky:
% Z-scaling and space--time structural relativity,
arXiv:hep-ph/0311306.

\bibit[5] G.S. Asanov, ``The Finsler-type recasting of Lorentz transformations."
In: Proceedings  of Conference  {\it Physical Interpretation of
Relativity Theory}, September 15-20 (London, Sunderland, 2000),
pp. 1-24.

\bibit[6] G.S. Asanov, \it Found. Phys. Lett. \bf15\rm, 199 (2002);
arXiv: gr-qc/0207089.

\bibit[7] G.S.~ Asanov: \it Rep. Math. Phys. \bf 45 \rm(2000), 155;
\bf 47 \rm(2001), 323; \bf 55 \rm(2005), 777.

%anisotr:
\bibit[8] G.S.~ Asanov: arXiv:gr-qc/0207117;
arXiv:hep-ph/0306023; arXiv:math-ph/0310019;
arXiv:math.MG/0402013.

\end{document}